\documentclass[twocolumn,aps,prc,superscriptaddress,showpacs]{revtex4}
\usepackage{amssymb}
\usepackage{amsmath,bm}
\usepackage{graphicx}
\usepackage{dcolumn}
\usepackage{multirow}
\usepackage{color}
\usepackage{CJK}
\usepackage{cases}
\begin{document}
\begin{CJK*}{GBK}{song}
\draft
\title{Multi-scaling mix and non-universality between population and facility density}

\author{J. H. Qian}
\affiliation{\footnotesize Shanghai Institute of Applied Physics,
Chinese Academy of Sciences, Shanghai 201800, China}
\author{ C. H. Yang}
\affiliation{\footnotesize School of Information Science and Technology, East China
Normal University, Shanghai 200241, China}
\author{ D. D. Han}
\affiliation{\footnotesize School of Information Science and
Technology, East China Normal University, Shanghai 200241, China}

\author{ Y. G. Ma}


\affiliation{\footnotesize Shanghai Institute of Applied Physics,
Chinese Academy of Sciences,  Shanghai 201800, China}


\date{\today}
\nopagebreak

\begin{abstract}

The distribution of facilities is closely related to our social
economic activities. Recent studies have reported a scaling relation
between population and facility density with the exponent depending
on the type of facility. In this paper, we show that generally this
exponent is not universal for a specific type of facility. Instead
by using Chinese data we find that it increases with Per Capital
GDP. Thus our observed scaling law is actually a mixture of some
multi-scaling relations. This result indicates that facilities may
change their public or commercial attributes according to the
outside environment. We argue that this phenomenon results from the
unbalanced regional economic level and suggest a modification for
previous model by introducing consuming capacity. The modified model
reproduces most of our observed properties.

\end{abstract}
\pacs{ 89.75.Hc, 89.75.Da, 89.40.Dd}

\maketitle

\section{Introduction}
  Facilities and infrastructures such as hospitals, schools, internet
routers, origin from the development of human social civilization
and in turn shape our modern daily life. This complex evolution
process raises an interesting question that how these facilities are
distributed and how they correlate with the whole social economic
system. A better understanding on this issue could help to provide
better public service and to save social opportunity cost. It is
believed that population density and economic ingredient are crucial
on deciding the locations of these facilities
\cite{internet,eco,jw,territorial,t2,political,newmanfr,newman}. But
unevenly distributed population and economic level make the question
very complicated. Although studies ranging from business economics,
system engineering, computer science, geography to even biology have
addressed on the issue, both theoretical basis and empirical
demonstration are inadequate
\cite{internet,eco,jw,territorial,t2,political,newmanfr,newman,bio1,bio2,bio3}.
This causes the arbitrary assumption of uniformly distributed nodes
in many spatial network models even though it is far from the
reality \cite{qian,spatial}.

  Intuitively, the number of facilities in an area increases with the
corresponding population. The related studies can be traced back to
a so-called \textit{p-median} problem which aims to find the precise
locations of facilities so that the mean distance that one reaches
his nearest facility is minimized \cite{pm,newman}. Numerical and
analytic treatments have been used to suggest a relation $D\sim
\rho^{\alpha}$ where $D$ is the facility density, $\rho$ is the
population density and $\alpha=2/3$
\cite{newmanfr,jw,newman,territorial}. This result accounts for the
internet router distribution and the territorial division but does
not consist with other empirical studies where although a universal
scaling $D\sim \rho^{\alpha}$ is evidenced, the value of $\alpha$,
depending on types of facility, ranges from $2/3$ to $1$ rather than
just fixed $2/3$ \cite{internet,territorial,pnas}. More
specifically, commercial facility and public one has $\alpha=1$ and
$\alpha=2/3$, respectively, while for facility with both attributes,
the exponent lies intermediate. These findings stimulated a recent
study in which a general model based on the tradeoff between the
profit  (commercial concern)  and the social opportunity cost
(public service) was proposed \cite{pnas}. The value of $\alpha$ in
that model represents the type of facility, which is described by
the relative weight of commercial and public attributes. In other
words it assumes a universal scaling exponent for a specific type of
facility. (Actually all the previous studies tacitly approve this
assumption.) Thus if we choose a part of these facilities according
to some other properties, the scaling exponent is expected to be
unchanged in this sample.

  As we will show in this paper, this is not always true. At least in
Chinese cases, samples from a specific type of facility according to
Per Capital GDP yield an increasing scaling exponent. Therefore the
macroscopic power law relation between facility and population
density is actually a mixture of different scaling functions. This
multi-scaling property indicates a different picture that the
attribute of a facility changes with the outside environment and
consequently affects its real distribution. In the next section we
will provide empirical evidences for the above arguments. And in
section III, we will try to offer an explanation and present a
modified model to reproduce the multi-scaling property.

\section{empirical study}
  We have gathered $7$-year empirical data ($2001-2007$) including the
positions of $4$ typical types of facility (hospital, post office,
school and theater), the population, GDP and the area of every
county. Despite of the temporal fluctuations, there are totally more
than $287$ counties and about $60500$ hospitals, $56300$ post
offices, $309200$ schools and $5000$ theaters. All these data come
from the CHINA CITY STATISTICAL YEARBOOK whose electronic versions
can be found and downloaded at http://ishare.iask.sina.com.cn. These
data allow us to calculate the population density, facility density,
Per Capital GDP and their correlation at county level. Such
coarse-grained treatment was also applied in the previous study due
to the resolution limitation of the facility positions
\cite{pnas,internet}.

\begin{figure}
\resizebox{21.4pc}{!}{\includegraphics{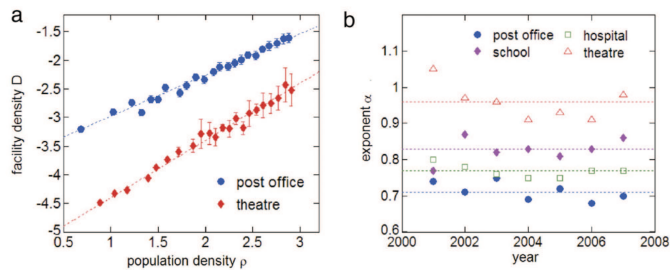}}
\caption{\footnotesize (a) The scaling law between facility density
($D$) and population density ($\rho$) for the cases of post office
and theater in $2007$. The data is logarithmic binned and is plotted
on the log-log scale. The dotted lines are the corresponding fits,
which are measured as $D\sim \rho^{0.7}$ (blue) and $D\sim
\rho^{0.98}$ (red). (b) The yearly scaling exponents for all the
four facilities. Dotted lines are their corresponding averages
calculated as $0.71, 0.96, 0.83, 0.77$ for post office, theater,
school, hospital, respectively. } \label{emone}
\end{figure}

  Our first observation consists with the previous studies. Indeed a
scaling relation between population and facility density emerges in
all the four types of facility (Fig.~\ref{emone}(a)). A detailed
analysis on their scaling exponents indicates that despite of the
yearly fluctuations, the exponents stay around their own averages
which are measured as $0.71, 0.96, 0.83, 0.77$ for post office,
theater, school, hospital respectively. According to Ref.~\cite{pnas},
post office and theater whose scaling exponents are close to $2/3$
and $1$ can be viewed as the representative of public and commercial
facility. This is consistent with our experience that post office
disregarding profit is necessary everywhere while theater behaves
oppositely.

  To examine the universality assumption of scaling exponent, we
classify the counties into four different classes. Specifically, we
calculate the Per Capital GDP for each county and divide their
logarithmic values into four equal intervals. Those counties whose
Per Capital GDP lie in a common interval belong to the same class.
Consequently counties in class $1$ have the lowest Per Capital GDP
while those in class $4$ have the highest one. This classification
distributes about $40$ counties in class $1$ and over $70$ ones in
each of the other three. Although the number of counties in class
$1$ is almost half less, the number of the corresponding facilities
is still larger enough to apply statistical analysis. Another
problem is whether the Per Capital GDP correlates obviously with the
population density so that our method might cause serious
statistical bias. This possibility is basically ruled out as the
correlation turns out to be very weak (correlation coefficient
$<0.25$).

\begin{figure}
\resizebox{20.2pc}{!}{\includegraphics{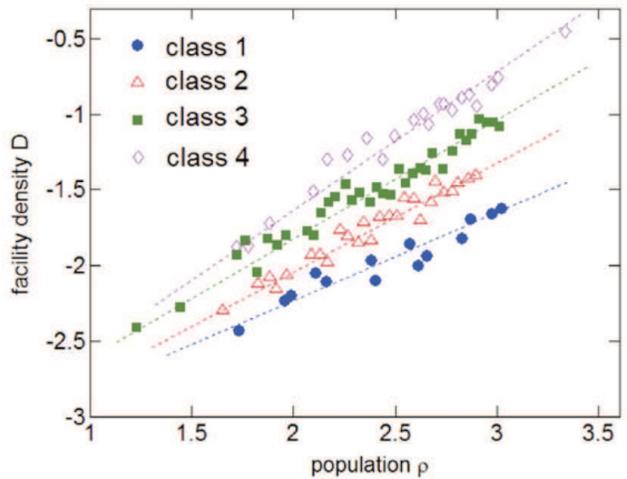}}
\caption{\footnotesize Multi-scaling relation for the case of
hospital in $2007$. The class from $1$ to $4$ describes the increase
of the Per Capital GDP level. For each class the data is logarithmic
binned and is plotted on the log-log scale. Dotted lines are their
corresponding fit plotted here for guiding  eyes. The data for
different classes is undrawn a little for better visualization. For
each class $D$ and $\rho$ still follows power law but the exponent
increases clearly with the class number. The result indicates a
multi-scaling picture rather than universality of the scaling
exponent.} \label{emtwo}
\end{figure}

\begin{figure}
\resizebox{21.2pc}{!}{\includegraphics{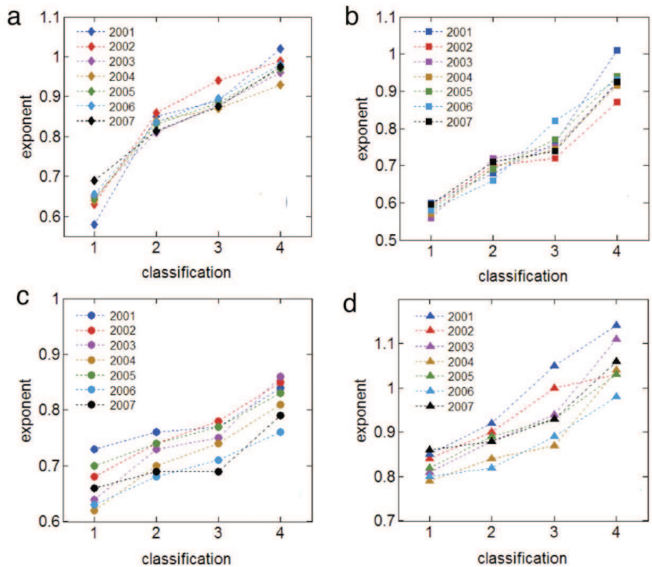}}
\caption{\footnotesize The relation between the multi-scaling
exponents (y-axis) and the Per Capital GDP level (class number or
classification as labeled on x-axis) for four types of facility: (a)
school, (b) hospital, (c) postoffice, (d) theater. For all the four
types of the facility the exponents increase with the Per Capital
GDP every year without exceptions. For school and hospital, the
exponents vary in a large range from $0.6$ to $1$. In contrast, the
exponents of post office and theater range narrowly.}
\label{emthree}
\end{figure}

  For each type of facility, we study their scaling relation in
different classes. As illustrated in Fig.~\ref{emtwo}, in each class
the relation between population and facility density is still a
power law. But their scaling exponents are not equal but increase
clearly with the Per Capital GDP level (class number). This result
contrasts with the universality assumption of the scaling exponent
in Ref. \cite{pnas}. Instead it indicates that the observed scaling
relation $D\sim \rho^\alpha$ is actually composed of some
multi-scaling behaviors. If we accept the physical meaning of
$\alpha$ interpreted in Ref.\cite{pnas}, the multi-scaling
phenomenon reveals an interesting fact that facilities can change
their attributes according to the outside economic environment.
Particularly, they tend to be commercial in high Per Capital GDP
level area but still provide necessary public services in poor
developed places. Further detailed analysis suggests that this
multi-scaling property as well as the positive correlation between
the multi-scaling exponents and Per Capital GDP level occurs every
year in all the types of facility regardless of their temporal
fluctuations (Fig.~\ref{emthree}). For school and hospital
(Fig.~\ref{emthree}(a) and Fig.~\ref{emthree}(b)), the exponents
vary from $2/3$ to $1$, which almost covers the whole possible
range. In contrast, the exponents of post office are much more
stable.  As shown in Fig.~\ref{emthree}(c) they stay near $0.7$ and
rise up to no more than $0.85$. Similarly in Fig.~\ref{emthree}(d),
the range of the exponents of theater narrows around $0.9$. It seems
that purely commercial or public facilities are not likely to behave
diverse attributes, which is somewhat consistent with our intuition.

\section{explanation and model modification}

  All these findings are not captured by the previous models and thus
require a more complete theory. As will be presented later, a small
modification by introducing the consuming capacity can reproduce
most of these properties. Before that, we will first review the
model of Ref.~\cite{pnas} and clarify some useful concepts.

  Suppose we are given the total population $N_p$ and the total number of
facilities $N_f$ on a plane. And suppose people always visit their
nearest facility, by which we can define the Voronoi cell $V_i$,
whose area is $s_i$, as the set of points closer to the $i$-th
facility than to any others\cite{newman}. This means that the number
of visitors to the $i$-th facility is the number of people living in
$V_i$, which is denoted as $n_i$. Therefore the population density
in $V_i$ is calculated as $\rho_i=n_i/s_i$ while the corresponding
facility density $D_i$ is $D_i=1/s_i$.

  For commercial facility, Ref.~\cite{pnas} assumes that the profit of the $i$-th
facility is proportional to $n_i$. Therefore, a facility having
lower $n_i$ is better off moved to other location with higher
population for higher profit. This strategy is applied by all the
facility during the relocating process. Finally the system will must
reach an equilibrium so that every facility has almost the same
profit, i.e. $n_i\sim N_p/N_f$. Then by using the expression
$\rho_i$ and $D_i$ calculated above, we arrive at $D\sim \rho$. On
the other hand, public facility concerns prior the social
opportunity cost caused by the distance between visitors and
facilities, which is described by $n_i \langle r_i \rangle$ with
$\langle r_i \rangle\sim \sqrt{s_i}$ representing the average
distance to the $i$-th facility. To provide better public service,
facilities at lower-cost places should be relocated to those with
higher $n_i \langle r_i \rangle$. Then in steady state, $n_i \langle
r_i \rangle$ becomes the same for all the facilities. Again by using
the expression $\rho_i$ and $D_i$, we have $D\sim \rho^{2/3}$. For
facilities with both attributes, Ref.~\cite{pnas} defines a general
quantity
\begin{equation}
c_i=n_i \langle r_i \rangle^{\beta},
\end{equation}
where $\beta$ is tunable within the range $[0,1]$. Analogous to the
above analysis, one can derive the final scaling relation $D\sim
\rho^{2/{(\beta+2)}}$, which gives $\alpha=2/{(\beta+2)}$.

  However in this nicely compact model, the assumption that profit
equals to population is  too simple. After all, the profit is
yet closely related to the commodity prices and people's consuming
capacity. Facilities providing luxury commodity are not opened at
poor places even though they are densely populated because no one
there can bear such high level consumption. On the other hand,
facilities in the regions with lower population but higher consuming
capacity can still benefit from high prices. Therefore facilities
with both attributes, due to their ingredient of commercial part,
take chance to locate and gain at developed area even though the
service there provided by their public ingredient is adequate
enough. This causes the number of facility in a well developed area
is larger than those in poor places even if they are equally
populated, which indicates a more rapid increase of facility in high
economic-level places. If we assume the consuming capacity has a
positive correlation with Per Capital GDP, which seems plausible,
the above explanation for multi-scaling does make sense.

  To make the explanation more convincing, it is useful to introduce
an alternative expression of Eq.(1) to characterize the transition
from public to commercial facility, expressed as
\begin{equation}
c_i=\lambda n_i+(1-\lambda)n_i \langle r_i \rangle,
\end{equation}
where $\lambda\in[0,1]$ is a tunable parameter controlling the
relative weight of commercial or public attribute and consequently
determining the final exponent $\alpha$, just as the role of $\beta$
in Eq.(1). Eq.(2) has the same physical meaning and similar effect to
Eq.(1) but turns out to be more difficult to apply analytical
treatment. However, one can still prove its scaling property by the
method used in Ref.\cite{epl}. To introduce the consuming capacity to
the model, we denote $m_i$ as the average expense consumed by every
person living in Voronoi cell $V_i$ so that the profit equals to
$m_in_i$. Then Eq.(2) is rewritten as
\begin{equation}
c_i=\lambda m_in_i+(1-\lambda)n_i \langle r_i \rangle.
\end{equation}
The only modification compared to Eq.(2) lies in the first term of
the righthand of Eq.(3), i.e. $\lambda\rightarrow \lambda m_i$. This
modification does not affect the macroscopic scaling law between $D$
and $\rho$ qualitatively, but changes the relative weight of
commercial attributes on microscopic level. Particularly, for large
$m_i$ the first term $\lambda m_in_i$ is enhanced by the effective
weight $\lambda m_i$, which causes the system to concern more about
profit. This leads the facility to be more commercial and the
scaling exponent to be close to $1$. On the other hand if $m_i$ is
small, the second term $(\lambda-1)n_i \langle r_i \rangle$ takes
over, then the system concerns more about social opportunity cost
just like public facility and the exponent becomes close to $2/3$.
Therefore even for the same $\lambda$ (i.e. the same type of
facility), different $m_i$ leads to different system behavior. This
is the reason why multi-scaling emerges and why their exponents
increase with the economic level. Moreover, if $\lambda=0$ (purely
public facility), Eq.(3) becomes $m_i$ independent. So the system
degenerates to the classical model in Ref.\cite{pnas} and thus
displays only a single scaling relation with the exponent
$\alpha=2/3$. On the other hand, if $\lambda=1$ (purely commercial
facility), Eq.(3) depends totally on the first term. But $m_i$ in
this case has no effect on the exponent but only changes the
coefficient of the scaling relation, leaving the only exponent
$\alpha=1$ which is also independent of $m_i$\cite{answer}.
Therefore multi-scaling property can be less pronounced in a very
commercial or public facility, which explains the narrow range of
the multi-scaling exponents observed in post office and theater.

\begin{figure}
\resizebox{20.8pc}{!}{\includegraphics{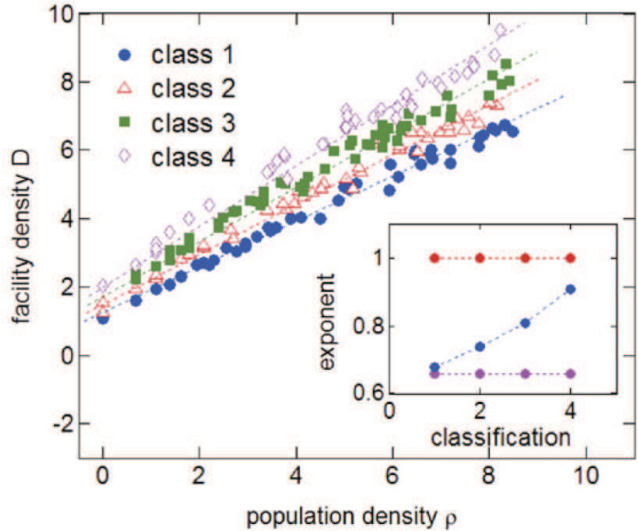}}
\caption{\footnotesize Multi-scaling relation between $D$ and $\rho$
simulated by the modified model with $\mu=1$. The data is plotted on
the log-log scale. The class from $1$ to $4$ describes the increase
of the level of $m$. The relation in each class is power law but the
exponent increases with the class number just as in
Fig.~\ref{emtwo}. Dotted lines are their corresponding fits plotted
here for guiding eyes. Inset: The simulated multi-scaling exponent
\textit{vs} class number. For $\mu=0$ (red) and $\mu\rightarrow
\infty$(purple), which represent the pure commercial and public
facility, the exponents stable at $1$ and $2/3$, respectively. For
intermediate $\mu$, such as $\mu=1$ (blue), the exponent increases
with the classification. The result is averaged over $50$
simulations.} \label{modelone}
\end{figure}
\begin{figure}
\resizebox{20.5pc}{!}{\includegraphics{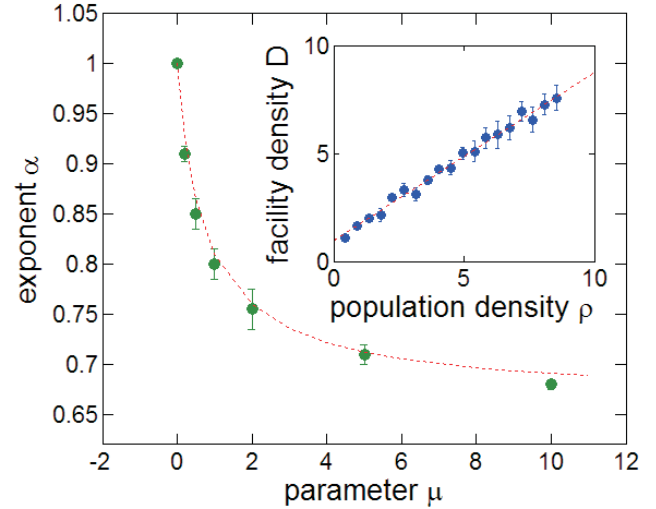}}
\caption{\footnotesize The relation between the scaling exponent
$\alpha$ and the parameter $\mu$. The green points are the
simulation result which is averaged over $50$ realizations. The red
dotted lines are the analytical prediction. Inset: The scaling
relation between population and facility density simulated by our
modified model with $\mu=1$. The simulation data is logarithmic
binned and is plotted on the log-log scale. Red dotted line is the
fit measured as $D\sim \rho^{0.8}$.} \label{modeltwo}
\end{figure}

If we still adopt the idea in Ref.\cite{pnas} and follow the
expression of Eq.(1), the above explanation indicates that the
exponent $\beta$ in Eq.(1) should be modified to be a function of
parameter $m_i$, i.e. $c_i=n_i \langle r_i \rangle^{\beta(m_i)}$. Then the
multi-scaling exponent can be calculated as $2/{(\beta(m_i)+2)}$ and
the macroscopic exponent $\alpha$ can be given by taking an average
over all possible $m_i$. To reproduce the multi-scaling property by
simulation, we apply the following simple formula of $\beta(m)$
\begin{equation}
\beta(m)=1-(m/m_{max})^{\mu},
\end{equation}
where $m_{max}$ represents the possible maximum of $m$ and $\mu$ is
a parameter controlling the sensitivity of $\beta$ with $m$. If
$\mu=0$, the facility becomes purely commercial while if
$\mu\rightarrow \infty$, it becomes purely public. So $\mu$ also
controls the exponent $\alpha$. The real-world $\beta(m)$ can be
quite different (probably related to the type of facility). Finding
its precise expression could be a complicated task and is beyond the
scope of this paper. The aim of our simulation is only to reproduce
the multi-scaling phenomenon qualitatively. Our simulation follows
the similar process to that of Ref.\cite{pnas} except for the
above-proposed modification. Specifically, we first distribute the
population density $\rho$ and the economic level $m$ randomly on a
plane. And then we further put some facilities of a certain type. At
each time step, every facility calculates its benefit according to
$c_i=n_i\langle r_i \rangle^{\beta(m_i)}$. Facility with the lowest $c_i$ then
moves to the region with the highest benefit. This process repeats
until the system is steady. At steady state, we measure various
quantities and their relations just as done for empirical data. Note
that the classification here is carried out according to the value
of $m$. Other details of the simulation such as parameter setting
are described in APPENDIX A. In Fig.~\ref{modelone}, we plot the
simulated relation between population and facility density for
different classes (i.e. different level of $m$ or say different
interval of $m$). Clearly it displays multi-scaling property with
the exponents increasing with class number. The inset of
Fig.~\ref{modelone} presents the results of the simulated
multi-scaling exponents. For purely commercial ($\mu=0$) or purely
public ($\mu\rightarrow \infty$) facility, the exponents become stable at
$1$ or $2/3$. But for intermediate $\mu=1$, the exponents increase.
In the inset of Fig.~\ref{modeltwo} we demonstrate that the whole
scaling relation between $D$ and $\rho$ maintains in our modified
model. And the scaling exponent $\alpha$ in the simulation decreases
with the parameter $\mu$, as shown in Fig.~\ref{modeltwo}. Note that
all these results can be calculated analytically. We present an
analytical solution of $\alpha(\mu)$ in Fig.~\ref{modeltwo}, which
is in good agreement with the simulation. More details about the
analytical calculations are reported in APPENDIX B.

\section{Conclusion}
  We have analyzed the scaling relation between population and
Chinese facility density at different Per Capital GDP levels. Our
study does not  see the universality of the scaling exponent but
instead suggests a multi-scaling picture. More interestingly, such
multi-scaling exponents increase with the Per Capital GDP regardless
of the type of facility. These results indicate that facilities can
change their commercial or public attributes according to the
outside environment, i.e. they take chance to gain more profit in
developed area but still fulfil their public-service responsibility
in poor region. We have also provided possible explanation and
suggest a modification by considering consuming capacity. The
modified model can reproduce most of our observed properties.

  Our study stresses on existence rather than universality. Indeed the
multi-scaling property is observed every year for all the four types
of facility. Therefore their occurrence is certain rather than
coincidental. On the other hand it is also appealing to explore
whether this phenomenon occurs in other more developed country where
the economic system is more balanced and stable. Due to the
limitation of data, the present study cannot cover this area. If
this point is evidenced, it indicates the multi-scaling could be a
common property. Otherwise it either requires explanations from
sociocultural, economic, politics, etc or leads us to a long-time
dynamic evolution picture of facility allocation, both of which are
significant on understanding our social economic system or even
providing guidelines to urban development.

This work was partially supported by the National Nature Science
Foundation of China under Grant Nos. 11075057, 11035009 and
10979074.

\appendix
\section{simulation details}
We use a coarse-grained simulation for our modified model. We first
divide a plane into many unit squares whose area all equal to $1$
and assume that all the situations in one unit square are
approximately identical. Then we distribute for each unit square $u$
the population $\rho_u$ and the consuming capacity $m_u$ according
to the corresponding distribution $p(\rho)$ and $p(m)$. And we
further distribute randomly the initial number of facility
$D_u(t=0)$. We use denotation $D_u(t)$ to emphasize that the
facility number changes with time during the simulation while
$\rho_u$ and $m_u$ are always fixed as soon as they are distributed.
Since the area of all $u$ is $1$, $\rho_u$ and $D_u$ is exactly the
population density and facility density respectively. The number of
people visiting facility $i$ in place $u$, denoted as $n_{iu}$, is
calculated as $n_{iu}=\rho_u/D_u$ while the average distance for
these people to travel to facility $i$, denoted as $\langle r_{iu}
\rangle$, is calculated as $\langle r_{iu}
\rangle\sim\sqrt{s_{iu}}=1/\sqrt{D_u}$\cite{answer2}. Then we can
determine the benefit $c_{iu}=n_{iu}\langle r_{iu}
\rangle^{\beta(m_u)}\sim\rho_u/{D_u^{1+\beta(m_u)/2}}$ for every
facility $i$ in the unit square $u$. Note that this quantity only
depends on place $u$ and is equal for any $i$ within this unit
square, so we can replace $c_{iu}$ by the denotation $c_u$. At each
time step of our simulation, we calculate the $c_u$ for each places
according to the current $D_u(t)$. Then we eliminate a facility in
the place with the lowest $c_u$ and create one in the unit square
with the highest $c_u$. This procedure repeats until the system
reaches its steady state, at which the relation between $D$ and
$\rho$ as well as the exponents is stable.

We test this coarse-grained method by repeating the simulation in
Ref.~\cite{pnas}, i.e. setting $\beta(m)=constant$ as Eq.(1). We find
the method behaves the same as the simulation in Ref.\cite{pnas} and
reproduces all their results. In our own simulation, we have $200$
different unit squares. And we set $p(\rho)=1/\rho$ with $\rho\in
[0,5000]$ and $p(m)=1/300$ with $m\in [0,300]$. The initial
distribution of facility is also uniform with $D(t=0)\in [0,1000]$.
We choose power-law distribution of $\rho$ because i). it coincides
with the real population distribution which is observed to be
heavy-tailed;  ii). the power-law formula leads to a uniformly
distributed data points on log-log plot, which gives a better
visualization. Note that other distributions do not change the
simulation results qualitatively. By these parameters, we obtain
$\beta(m)=1-(m/300)^{\mu}$.

When the simulation reaches its steady state, we measure various
quantities and their relations just as done for empirical data. Note
that the classification here is carried out according to the value
of $m$. Specifically, we divide the value of $m$ (not the logarithmic
value of $m$) into four equal intervals. Those unit square $u$ whose
$m_u$ lies in a common interval belongs to the same class. Although
there is a bit difference from what we have done for empirical data,
it does not change our conclusion at all.

\section{analytical calculations}
Since
$\alpha=\frac{d \langle\ln(D_i)\rangle_{m_i}}{d\ln(\rho)}={\langle \alpha_i\rangle_{m_i}}$,
the macroscopic exponent is exactly an average of $\alpha_i$ over
all possible $m_i$, i.e.
$\alpha=\int_{m_{min}}^{m_{max}}p(m)\alpha(m)dm$, where
$\alpha(m)=2/{(\beta(m)+2)}$. Substituting the corresponding
parameters and equation, we have
\begin{equation}
\alpha(\mu)=\int_{0}^{1}\frac{2}{3-x^\mu}dx.
\end{equation}
This function gives a good agreement with the simulation, as plotted
in Fig.~\ref{modeltwo}. Particularly, for $\mu=0, 0.2, 0.5, 1, 2, 5, 10$,
Eq.(B1) gives $\alpha=1, 0.926, 0.866, 0.81, 0.76, 0.712, 0.692$, which is
consistent with the simulation data
$\alpha=1, 0.91, 0.85, 0.8, 0.755, 0.71, 0.68$ in Fig.~\ref{modeltwo}. We
can also calculate the multi-scaling exponent for each class. The
calculation is given by
$\alpha(C)=\int_{m_{min}}^{m_{max}}p(m|C)\alpha(m)dm$, where
$p(m|C)$ is the conditional probability density under class $C$ and
equals to $1/75$ in our simulation. $m_{min}$ and $m_{max}$ here is
related to the specific class $C$. We calculate the multi-scaling
exponents for $\mu=1$ and have the theoretical result
$0.69, 0.76, 0.84, 0.94$ for class $1, 2, 3, 4$, which is also consistent
with the simulation data $0.68, 0.74, 0.81, 0.91$ in the inset of
Fig.~\ref{modelone}.

 {}
\end{CJK*}

\begin{thebibliography}{}
\bibitem{answer} In this case the scaling relation for a specific
$m_i$ is $D=m_i\frac{N_f}{\langle m \rangle N_p}\rho$. Therefore the scaling
exponent $\alpha=1$ regardless of $m_i$.
\bibitem{answer2} Since we assume that all the situations in one unit square are
identical, both the number of people and the area belonging to a
facility are equal everywhere, leading $n_i$ and $s_i$ to equaling
to their corresponding average, i.e. $\rho_u/D_u$ and $1/D_u$,
respectively.
\end{thebibliography}
\end{document}